\documentclass[information,article,moreauthors,pdftex,10pt,a4paper]{mdpi} 
\firstpage{1} 
\history{Received: 14 February 2018; Accepted: 27 March 2018; Published: date}

\pdfoutput=1
\usepackage{color}
\preto{\abstractkeywords}{\nolinenumbers}
\Title{Random Linear Network Coding for 5G Mobile Video Delivery}

\Author{Dejan Vukobratovic $^{1,}$*, Andrea Tassi $^{2}$, Savo Delic $^{1}$ and Chadi Khirallah $^{3}$}

\AuthorNames{Dejan Vukobratovic, Andrea Tassi, Savo Delic and Chadi Khirallah}

\address{%
$^{1}$ \quad Faculty of Technical Sciences, University of Novi Sad, 21000 Novi Sad, Serbia; savodelic@gmail.com\\
$^{2}$ \quad Department of Electronic and Electrical Engineering, University of Bristol, Bristol BS8 1TH, UK; a.tassi@bristol.ac.uk\\
$^{3}$ \quad School of Engineering, University of Edinburgh,  Edinburgh EH8 9YL, UK; C.Khirallah@ed.ac.uk} 

\corres{Correspondence: dejanv@uns.ac.rs; Tel.: +381-21-485-2535}

\abstract{An exponential increase in mobile video delivery will continue with the demand for higher resolution, multi-view and large-scale multicast video services. Novel fifth generation (5G) 3GPP New Radio (NR) standard will bring a number of new opportunities for optimizing video delivery across both 5G core and radio access networks. One of the promising approaches for video quality adaptation, throughput enhancement and erasure protection is the use of packet-level random linear network coding (RLNC). In this review paper, we discuss the integration of RLNC into the 5G NR standard, building upon the ideas and opportunities identified in 4G LTE. We explicitly identify and discuss in detail novel 5G NR features that provide support for RLNC-based video delivery in 5G, thus pointing out to the promising avenues for future research.}


\begin{document}

\section{Introduction}

Mobile video delivery continues its growth in volume and will reach an estimated $78\%$ of the total mobile data traffic by 2021, compared to $60\%$ in 2016 \cite{VNI}. In absolute values, during the same period (2016--2021), the total volume of mobile data traffic will experience seven-fold increase \cite{VNI}. This increase is due to the combination of ever increasing resolutions of user handsets and proliferation of 4K/8K ultra high-definition (UHD) formats \cite{UHD}, fueled by evolution of innovative video services relying on multi-view and 360-degree video \cite{5GVR}, enhanced broadcast \cite{BCASTMM} and peer-to-peer video services \cite{P2PMM}.

Evolution of mobile cellular infrastructure capable to cope with the surge of video traffic is necessity for meeting these predictions. Towards this end, 3rd Generation Partnership Project (3GPP) have just completed the first phase in defining a new fifth generation (5G) New Radio (NR) interface~\cite{5GNR}. One of the resting pillars of 3GPP 5G NR is the support for enhanced Mobile Broadband (eMBB) services that target providing users with sufficient data rates to accommodate new, high-rate mobile video services. The support for eMBB will be achieved by novel throughput-enhancement solutions implemented both in radio access network (RAN) \cite{5GRAN}, such as massive multiple-input multiple-output (MIMO) antenna technology or migration to wider millimeter-wave bands (mmWave), and in core network (CN) domain, by supporting network slicing via network function virtualization (NFV) and software-defined networking (SDN) technologies \cite{5GCN}.

For mobile video multicast/broadcast services, mobile cellular networks provide support in the form of 3GPP-defined enhanced Multimedia Broadcast/Multicast Service (eMBMS) \cite{eMBMS}. However, majority of mobile video traffic represents over-the-top (OTT) video streaming such as progressive downloading (PD) and adaptive bitrate streaming (ABR) for which mobile cellular network protocols remain largely oblivious \cite{QoEVoLTE}. In order to optimize mobile video delivery, a number of solutions have been proposed in recent research studies, including video-aware resource allocation \cite{QoESch1} and proactive edge caching and processing \citep{EdgeCache,LontheEdge,MEC}.

In this  review paper, we focus on random linear network coding (RLNC) as a scheme identified to create potentially high impact on flexible, efficient and reliable mobile video delivery. RLNC is a packet-level erasure protection mechanism that is simple, efficient and has a number of useful features including rateless property, i.e., capability to produce arbitrary many encoded packets from a given source block, and network coding property, i.e., capability to increase throughput in certain network scenarios by re-encoding packets in intermediate network nodes \citep{HetAl2006,FBW2006,CW2007,FSbook2007}. We provide an overview of RLNC placed in the context of a packet-level data processing protocol sublayer that can be easily integrated at different layers of protocol stack within mobile video delivery environment. The RLNC sublayer can be further optimized and improved with respect to complexity and video quality using sparse and unequal error protection RLNC design \citep{TCL2016,BJT2018,Feizi2014,VS2012}. In parallel with the review of RLNC, we provide an in-depth overview of mobile video delivery, focusing on unicast and multicast/broadcast mobile video services over fourth generation (4G) Long-Term Evolution (LTE) network \citep{svccomm,PD,ABR,eMBMS}. We then move on to investigate how RLNC sublayer can be integrated as part of the 4G LTE mobile video delivery services, discussing various options for both video unicast and multicast/broadcast across different protocol layers. In the final part, we provide possible directions for optimizing mobile video delivery and integrating RLNC sublayer within upcoming 5G NR standard. We conclude this overview paper by explicitly identifying and discussing in detail novel 5G NR features that could provide support for RLNC-based video delivery in 5G, thus pointing out to the promising avenues for future research.

The paper is organized as follows. In Section \ref{sec2}, we provide an overview of RLNC, presenting it as a basis of a modular RLNC-based protocol sublayer. We present performance measures and possible design extensions of the RLNC sublayer. In Section \ref{sec3}, we review in detail video delivery over 4G LTE mobile cellular networks, focusing on two main types of services: OTT video unicast and eMBMS-based video multicast/broadcast services. We present a review of practical mechanisms and academic investigations, targeting both CN and RAN design, that aim to support and enhance 4G LTE mobile video delivery. In Section \ref{sec4}, we discuss integration of RLNC sublayer across different layers of 4G LTE protocol stack. Based on these insights, in Section \ref{sec5}, we identify novel opportunities and challenges for RLNC sublayer and, in general, for mobile video delivery optimization, within 3GPP standardized 5G NR. The paper is concluded in Section \ref{sec6}.  
 
\section{Overview of Random Linear Network Coding \label{sec2}}

In this section, we provide a generic overview of a packet-level RLNC method for erasure protection across packet erasure channels in both the unicast and the multicast/broadcast scenario. We adopt a modular approach where RLNC is set as a core component of a RLNC protocol sublayer, whose integration in mobile video delivery solutions will be discussed in the rest of the paper. For more detailed account on the theory of RLNC, we refer interested reader to \citep{HetAl2006,FBW2006,CW2007,FSbook2007}.
 
\subsection{Introduction to RLNC}

\begin{figure}[H]
	\centering
	\includegraphics[width=15.0cm]{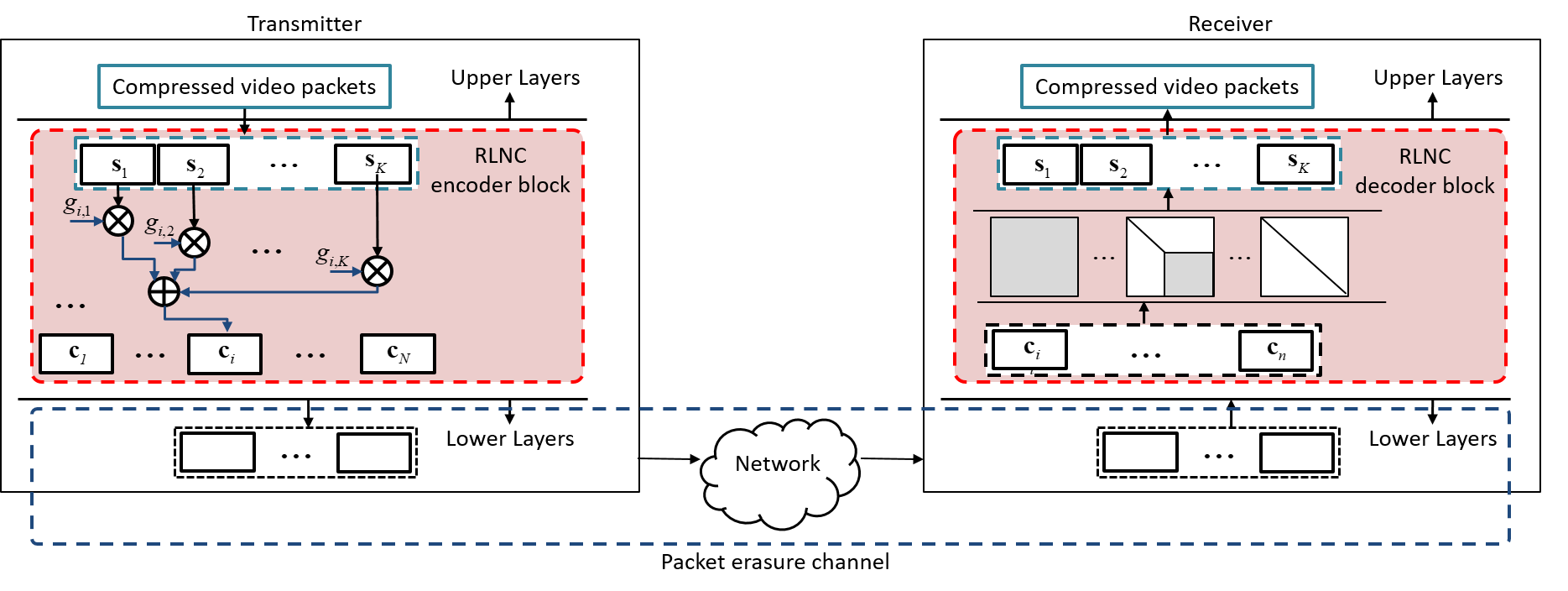}
	\caption{Generic RLNC sub-layers at the transmitter and the receiver side connected via packet erasure~channel.}
	\label{Fig_1}
	\end{figure}

The system model under consideration contains RLNC encoder block at the transmitter and RLNC decoder block at one or more receivers, connected via independent packet erasure channels. An~example with a single transmitter and a single receiver is illustrated in Figure \ref{Fig_1}.

\textbf{RLNC encoder block}: The input to the RLNC encoder is a source block $\mathbf{s}=[\mathbf{s_1}, \mathbf{s}_2, \ldots, \mathbf{s}_K]$ containing $K$ equal-length \emph{source packets}, each containing $L$ symbols of a finite field $\mathbb{F}_q$ of size $q$. RLNC encoder encodes $\mathbf{s}$ into a stream of \emph{coded packets} $\mathbf{c}=[\mathbf{c}_1, \mathbf{c}_2, \ldots, \mathbf{c}_N]$, where each encoded packet $\mathbf{c}_i$ represents a random linear combination of the source packets, i.e., $\mathbf{c}_i=\sum_{j=1}^{K} g_{i,j}\mathbf{s}_j$, and is of the same length $L$ symbols of $\mathbb{F}_q$ as source packets. The \emph{coding coefficients} $g_{i,j}\in \mathbb{F}_q$ are selected uniformly at random from $\mathbb{F}_q$, and for each $\mathbf{c}_i$, the associated set of coding coefficients forms the \emph{coding vector} $\mathbf{g}_i=[g_{i,1}, g_{i,2}, \ldots, g_{i,K}]$. Note that the transmitter can produce arbitrarily many encoded packets $N$ from $K$ source packets in a rateless fashion. Fixing $N$, the set of encoded packets can be represented as $\mathbf{c}=\mathbf{s}\cdot \mathbf{G}^T$, where \emph{coding matrix} $\mathbf{G}$ represents a $N \times K$ random matrix over $\mathbb{F}_q$. Frequently, systematic RLNC is also considered, where $\mathbf{c}_i=\mathbf{s}_i, 1 \leq i \leq K,$ i.e., the first $K$ coded packets are replicas of source packets, thus the first $K$ rows of $\mathbf{G}$ represent $\mathbf{I}_{K \times K}$ identity matrix.

\textbf{Packet erasure channel}: Coded packets are transmitted to one or more receivers via independent packet erasure channels. For the $j$-th receiver, the erasure probability of the corresponding channel is denoted as $\epsilon_j,$ where $0 \leq \epsilon_j \leq 1$. If the focus is on a single receiver, we will omit index and use $\epsilon$ as erasure probability.   Packet erasure channel is adopted as, further in the paper, we will only consider integration of RLNC as part of the higher protocol layers, without considering any details or information available from physical (PHY) layer.

\textbf{RLNC decoder block}: Due to possible packet losses, a receiver receives a subset of $n \leq N$ coded packets. Extracting the corresponding coding vectors, the receiver obtains $\mathbf{c}=\mathbf{s}\cdot \mathbf{D}^T$, where, with some abuse of notation, $\mathbf{c}=[\mathbf{c}_1, \mathbf{c}_2, \ldots, \mathbf{c}_n]$ represents the set of received coded packets, while $\mathbf{D}$ is a random $n \times K$ matrix over $\mathbb{F}_q$. To recover source packets, the receiver applies Gaussian Elimination (GE) decoding, which successfully recovers the source block $\mathbf{s}$ iff for the rank $r(\mathbf{D})$ of the decoding matrix $\mathbf{D}$ it holds that $r(\mathbf{D}) = K$.  We note that for shorter source block lengths $K$, the complexity of GE decoding is acceptable, which is why longer source blocks are typically divided into sub-blocks called generations \cite{CW2007}. The complexity of GE decoding can be further reduced by exploiting systematic, sparse and tunable sparse RLNC we describe shortly below.

\textbf{Performance measures}: One of the main RLNC performance measures is the probability the source block $\mathbf{s}$ is successfully recovered given the number of received coded packets $n$. This \emph{decoding probability} can be easily calculated \cite{C2000,TC2011}:
\begin{equation}\label{eq1}
P_d(n)=\left\{
\begin{array}{ll}
0 & \text{if} \: n<K, \\
\prod_{i=1}^{K-1}(1-\frac{1}{q^{n-j}}) & \text{if} \: n \geq K.
\end{array} \right.
\end{equation}

Note that \eqref{eq1} represents a cumulative distribution function (cdf) of the probability that $K$ linearly independent packets are collected among $n$ received coded packets. The corresponding probability density function (pdf) is $p_d(n)=P_d(n)-P_d(n-1)$. 

It is often useful to take the transmitter perspective and introduce a time reference assuming that each coded packet transmission takes a unit-time slot. If we fix the number of transmitted coded packets to $N \geq K$, we can (re)define the {decoding probability} after $N$ coded packets are transmitted:
\begin{equation}\label{eq2}
P_d(N)=\sum_{n=K}^{N} {N \choose n} \epsilon^{N-n} (1-\epsilon)^{n} P_d(n).
\end{equation}
Several interesting performance measures immediately follow. \emph{Outage probability} $P_o(N)$ is a probability the receiver will not recover the source block after $N$ transmitted coded packets: $P_o(N)=1-P_d(N).$ Average number of coded packet transmissions $\bar{N}$ required for successful source block decoding, also referred to as \emph{average decoding delay}, can be investigated for fixed $N$, and for the case $N \rightarrow \infty$~\cite{Nistor2011}. For~the former case, closed-form expressions for average decoding delay are available, while for the latter case, the upper bounds have been derived, both for the systematic and non-systematic RLNC~\cite{CT2017}.    

\subsection{Sparse RLNC}

Standard RLNC described above is limited by the decoding complexity of the GE decoder that scales as $O(K^3)$ with the source block size. To some extent, and under low erasure rates, systematic RLNC approach may alleviate the complexity issue by removing received systematic packets from the decoding process. More flexible solutions resort to sparse RLNC (S-RLNC), where sparse random linear combinations are used to generate coded packets. This is typically done by changing the random sampling process of coding coefficients by promoting zero-valued coefficients:
\begin{equation}
\mathbb{P}(g_{i,j}=v)=\left\{
\begin{array}{ll}
t & \text{if} \: v=0, \\
\frac{1-t}{q-1} & \text{if} \: v \in \mathbb{F}_q \setminus \{0\}.
\end{array} \right.
\end{equation}
Similarly as in RLNC, S-RLNC schemes are also investigated for decoding probability, outage probability and average decoding delays. However, even for $P_d(N),$ exact expressions are unknown but only approximated by upper/lower bounds \cite{TCL2016,BJT2018}.

Adaptive extension to S-RLNC is proposed in the form of Tunable Sparse RLNC (TS-RLNC) \cite{Feizi2014}. In TS-RLNC, at the beginning of a session, sparse linear combinations are used, while as the session continues, the coding density is increased, which may be tuned for desired trade-off between decoding complexity and average decoding delay.       

\subsection{Unequal Error Protection RLNC}\label{sec.UEP-RLNC}

In typical video delivery scenarios, the source block $\mathbf{s}$ can be divided into $L$ sub-blocks or \emph{layers} such that $\mathbf{s}=[\mathbf{l_1}, \mathbf{l}_2, \ldots, \mathbf{l}_L]$, where the $i$-th layer $\mathbf{l}_i$ contains $k_i$ source packets and $\sum_{i=1}^L k_i=K$. As the layer index $i$ grows, the packets contained in $\mathbf{l}_i$ have progressively decreasing impact on reconstructed video quality. For this scenario, unequal error protection RLNC (UEP RLNC) schemes offer significant benefits in terms of flexibility, reconstructed video quality and decoding complexity, as compared to the standard RLNC.

Two generic and well-studied UEP RLNC methods are non-overlapping window RLNC (NOW-RLNC) and expanding window RLNC (EW-RLNC) \cite{VS2012}. In both schemes, a set of $L$ windows $\mathbf{w}=[\mathbf{w}_1, \mathbf{w}_2, \ldots, \mathbf{w}_L]$ is defined over the set of layers: for NOW-RLNC, windows correspond to layers, i.e., $\mathbf{w}_i=\mathbf{l}_i$, while for EW-RLNC, the $i$-th window $\mathbf{w}_i$ contains the first $i$ layers, i.e., $\mathbf{w}_i=[\mathbf{l}_1, \mathbf{l}_2, \ldots, \mathbf{l}_i]$. Coded packets are produced by applying RLNC over a content of a window randomly selected using a window selection distribution $\mathcal{W}$. Proper design of $\mathcal{W}$ achieves desired balance of decoding probabilities of source packets belonging to different layers \cite{VS2012}.  

\subsection{RLNC Extensions to Erasure Networks}

In this paper, we restrict our attention to single-hop erasure channels, either in unicast or multicast/broadcast scenarios. Such models will be sufficient for our RLNC-based mobile video delivery considerations later in Sections IV and V. Before proceeding, we make two remarks. 

\textbf{RLNC and Rateless Codes:} For single-hop channels considered in this paper, RLNC schemes described above represent instances of rateless codes \cite{LPC2010}. Thus one can replace RLNC with other popular classes of rateless codes such as LT codes \cite{LT2002} or Raptor codes \cite{Raptor2006}. Indeed, Raptor codes provide near-optimal performance under significantly lower decoding complexity, thus allowing for larger source block lengths. The main benefit of using RLNC is that their usage is easily extended to multi-hop erasure network models where coding is performed in intermediate nodes. On the other hand, the source blocks lengths $K$ in video delivery scenarios are typically small, thus using RLNC usually does not incur significant decoding complexity penalty.

\textbf{Extensions to Erasure Network Models:} RLNC emerged as a practical solution to the network coding problem, where RLNC is applied in intermediate nodes of erasure network models \cite{CW2007}. Since then, RLNC has been investigated in various erasure networks scenarios. Among these, we point out to line networks \cite{PF2005}, and more general multicast and multiple-unicast models \cite{LMK2006,SMR2011}, as the models of interest for RLNC applications in future dense mobile cellular networks.

\section{Overview of Video Delivery in 4G Mobile Cellular Networks \label{sec3}}

This section reviews mobile video delivery in 4G LTE mobile cellular networks. We first provide background information on standard video content formats, unicast, and multicast/broadcast services in LTE. Then we provide specific details on LTE CN and RAN support for mobile video delivery.

\subsection{Mobile Video Delivery in 4G LTE}
%
%
\textbf{Video Coding Standards:} Video codecs are under constant evolution due to ever increasing performance requirements and novel use cases. The current video coding standards, H.265/HEVC~\cite{h265hevc}, replaced the previous one, H.264/AVC \cite{h264avc}, due to requirements for higher coding efficiency, higher spatial resolution (4K/8K video), color resolution and dynamic range. Extensions of HEVC include scalable (SHVC), multi-view (MV-HEVC), range (RExt) and 3D video coding (3D-HEVC) \cite{hevcExt}.

Details of HEVC compression are beyond the scope of this paper. We assume compressed video is packetized and organized into source blocks compatible with RLNC coding approach in Section \ref{sec2}. Typically, source blocks represent compressed group of frames (GOFs). Layered source block structure can be obtained via scalable video coding (SVC) or using specific codec features such as slicing or data partitioning \cite{svccomm,Nazir2015}. Finally, we note that, besides H.264/AVC and H.265/HEVC, other popular video codecs are in use such as VP9 and AV1 codecs.

\textbf{Mobile Video Streaming/Downloading over 4G LTE:} Most prevalent techniques for online video delivery are progressive downloading (PD) \cite{PD} and adaptive bitrate streaming (ABR) \cite{ABR}. Chronologically, progressive downloading was first implemented and aimed to enable video users watching the video before the entire video content is downloaded. Video players download first metadata which describes video details and as soon as the first video data has been downloaded the rendering can start. ABR streaming also provides users capability to watch video content before the download is complete, however, this streaming technique provides multiple representations of the same video on the content server. These representations are encoded in different resolutions and bitrates thus allowing video clients to adapt delivered video resolution by switching between different representations according to the bandwidth available on the client side. Multiple ABR implementations are available but the most dominant ones are Apple HLS (HTTP Live Streaming), DASH (Dynamic adaptive streaming over HTTP), Microsoft Smooth Streaming and Adobe HTTP Dynamic streaming. Analysis shows that around 80$\%$ of total mobile streamed video is delivered in ABR format while the rest is delivered in PD format. 70$\%$ out of total mobile streamed video is delivered in encrypted format by using HTTPS or QUIC transport, where e.g., QUIC protocol is used for Youtube content.

\textbf{Mobile Video Multicasting/Broadcasting over 4G LTE:} LTE network support for a point-to-multipoint (PtM) services is defined in 3GPP as evolved Multimedia Broadcast/Multicast Service (eMBMS) \cite{eMBMS}. The service is initially designed for mobile TV and radio broadcasting use case, however, other push-based services such as popular content caching (e.g., podcasts, news, ads, updates), live streaming of popular events (e.g., sport events such as olympics) and mobile network emergency alerts contributed to eMBMS development. eMBMS is delivered via PtM radio bearers thus providing for efficient usage of radio resources for the price of using fixed (i.e., non-adaptive) and conservative transmission configuration targeting improved cell coverage. In the context of this paper, eMBMS provides support for application-layer forward error correction (AL-FEC) \cite{CMGG2013}, where Raptor codes are recommended, although potentially, RLNC could also be used as an alternative. Despite rising interest in LTE broadcasting, eMBMS has not yet been massively deployed at mobile network operators, while significant experience and promising prospects are gained via service trials \cite{LTE-B-WP}.

\subsection{Core Network Support for Mobile Video Delivery over 4G LTE \label{sec3.2}}

\begin{figure}[H]
	\centering
	\includegraphics[width=15.0cm]{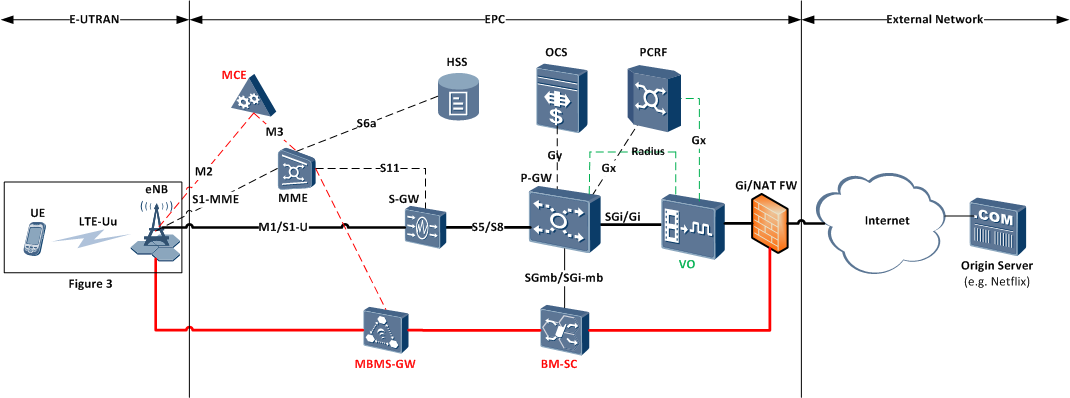}
	\caption{EPC network elements supporting mobile video delivery over 4G LTE.}
	\label{Fig_2}
	\end{figure}

\textbf{4G LTE Network Architecture:} Figure \ref{Fig_2} illustrates LTE network architecture that consists of two main parts: (i) evolved universal terrestrial radio access network (E-UTRAN); and (ii) evolved packet core (EPC). Due to functional split, EPC separates user plane and control plane elements. User plane elements,  Serving-Gateway (S-GW) and Packet Data Network (PDN)-Gateways (P-GW), provide data connectivity between E-UTRAN and external PDN. The S-GW handles the user-plane packet data termination towards E-UTRAN, while P-GW interfaces with the external PDNs performing IP related functions such as IP address allocation, policy enforcement, packet classification and routing. The main control plane element, Mobility Management Entity (MME), is responsible for connection/release of radio bearers to user equipment (UE). Further control plane entities include the Policy and Charging Rules Function (PCRF), enforcing policies and rules that are configured statically or dynamically per subscriber data session, Home Subscriber Server (HSS) that contains subscriber data such as user/QoS/barring profiles, and Online Charging System (OCS) used for real time rating of subscriber data usage and providing subscribers with data usage control. For more details on EPC architecture and elements, we refer interested reader to \cite{TS23.002,DPBook}.   

Popular over-the-top (OTT) mobile video unicasting services such as ABR or PD do not require additional EPC elements, as they are transmitted transparently over the EPC data bearers. However, as described next, due to massive volume of ABR/PD traffic mobile operators often empower their EPC with video optimization (VO) platforms. In contrast, for video multicasting/broadcasting, significant EPC upgrade is needed to provide eMBMS, as presented in Figure \ref{Fig_2}.

 \noindent

\textbf{Core Network Support for Video Streaming/Downloading:} To provide support for optimized unicast video delivery, mobile operators introduce VO platforms (Figure \ref{Fig_2}). VO platform is used to improve subscriber experience by optimizing video content to a format which will provide the best viewing experience for the given subscriber’s network conditions. The VO platform consists of traffic classifier (TC) and the system that performs optimization of unicast streams (VO-subsystem). The video traffic, which is sent by content server towards a content consumer (mobile subscriber), is intercepted by the VO platform, where it is first detected by TC, passed to the VO-subsystem, optimized, and sent towards the content consumer. The most frequently used VO optimization methods are:
\begin{itemize}
\item \textbf{Transcoding:} converts video content from one format to another by changing e.g. encoding format, resolution, codecs, frame rate, etc. Online (on-the-fly) transcoding is mostly used. In addition, offline transcoding can be used and it is done in a way that some popular videos are downloaded and optimized in advance before being stored in cache \cite{svccomm,PD}.
\item \textbf{Transrating:} converts video by keeping the original video format and resolution and by changing number of bits per pixel. This technique is not widely used \cite{PV2012}.
\item \textbf{ABR pacing:} receives traffic from a content server with one pace and send it towards a consumer with another pace in order to limit the representation quality requested by ABR clients on the subscriber side. This technique changes effective bandwidth perceived by ABR client side in order to affect the representation quality that will be selected by the client \cite{PV2012,E2015}.
\item \textbf{JIT (just-in-time) pacing:} receives traffic from a content server with one pace and lowers the downstream pace in ``just in time'' manner. This is done in order to avoid unnecessary filling of video player buffer on subscriber side as well as waste of network resources \cite{PD}.
\item \textbf{ABR manifest file manipulation:} consists of interception of ABR manifest file at the VO platform, parsing it and filtering out the representations from the manifest file that are not possible to reproduce on subscriber side under current network conditions.
\item \textbf{Caching (transparent and selective):} consists of storing popular traffic on the platform in order to make it available for future subscriber’s requests \cite{Caching,Gorilla}. Transparent caching strategy consists of caching of all unencrypted content. As it is resource consuming operators typically uses selective caching such as caching of the content that is previously transcoded.
\end{itemize}

Table \ref{Table_1} provides the applicability of the above-listed techniques on different video traffic types. Whether the traffic is optimized by VO-subsystem depends on its profile configuration, which may depend on: (i) time of the day (e.g., during peak hours more aggressive optimization can be done); (ii) radio access technology (RAT) type of the subscriber (e.g., allowed max bitrate for 4G RAT type is higher than for 3G RAT type) which is received from P-GW via Radius interface (Figure \ref{Fig_2}); (iii)~assigned subscriber policy received from PCRF via Gx interface (Figure \ref{Fig_2}); (iv) RAN congestion state (e.g., if RAN is in congested state then more aggressive optimization can be used for all subscribers to mitigate congestion); (v) location of subscriber (e.g., if the subscriber is associated to congested cell then more aggressive optimization is applied), and vi) device class (e.g., streams destined to devices with small screens can be optimized in more aggressive way than streams to devices with a wide~screens). 

\begin{table}[H]
\centering \caption{VO optimization methods and traffic types.}
\label{Table_1}
\begin{tabular}{cc}
\toprule
\textbf{Video Traffic Type} & \textbf{Possible Optimization Method}\\
\midrule
ABR over HTTP & ABR pacing; JIT pacing; ABR manifest file manipulation \\ 
ABR over HTTPS & ABR pacing; JIT pacing;\\ 
ABR over DRM over HTTP & ABR pacing; JIT pacing;\\
ABR over QUIC & ABR pacing; JIT pacing;\\
PD over HTTP & JIT pacing; online/offline transcoding; transrating; caching \\ 
PD over HTTPS & Not possible to optimize \\ \bottomrule
\end{tabular}
\end{table}

ABR over HTTPS is dominant video format, thus making ABR and JIT pacing the most commonly used optimization methods in VO platforms. We note that in academic studies, video quality-based optimization methods for ABR video delivery is currently very popular research topic \cite{E2015,Medieval}.

\textbf{Core Network Support for Video Multicasting/Broadcasting:} The entry point for eMBMS video content is broadcast/multicast service centre (BM-SC), which schedules and announces eMBMS services to end users. BM-SC is where AL-FEC is applied for packet-level erasure protection, which can be further optimized for eMBMS live streaming services \cite{CMGG2013,ALFECeMBMS}. Based on the content and control inputs from BM-SC, MBMS Gateway (MBMS-GW) establishes IP multicast session towards all eNBs that deliver eMBMS service to end users, supported by control signaling via MME. Single frequency network service (MBSFN) is commonly used, where the content is delivered with tight synchronization requirements enforcing identical physical layer (PHY) configuration at all eNBs to enhance reception at cell edges. Alternatively, single-cell point-to-multipoint service (SCPTM) can be used targeting eMBMS service in a specific cell \cite{eMBMS}. An overview of network deployment options for both video unicast and eMBMS services over LTE are provided in \cite{QoEVoLTE}.   

As depicted in Figure \ref{Fig_2}, the data path (solid red line) traverses BM-SC, MBMS-GW and proceeds via PTM IP Multicast session to the set of eNBs in MBSFN service area in the case of MBSFN service, or directly to a specific eNB in case of SCPTM service. The supporting signaling (dashed red line) traverses MME and Multi-cell/Multicast Coordinating Entity (MCE), where MME is responsible for session-level management, while MCE controls configuration of eNBs in MBSFN area, such as coding and modulation, time synchronization and resource allocation \cite{eMBMS}. Overall, providing eMBMS services requires significant CN upgrade, which, together with lack of killer applications \cite{LTE-B-WP}, contributes for slow deployment of eMBMS at mobile operators.      

\subsection{Radio Access Network Support for Mobile Video Delivery over LTE}

\begin{figure}[H]
	\centering
	\includegraphics[width=15.0cm]{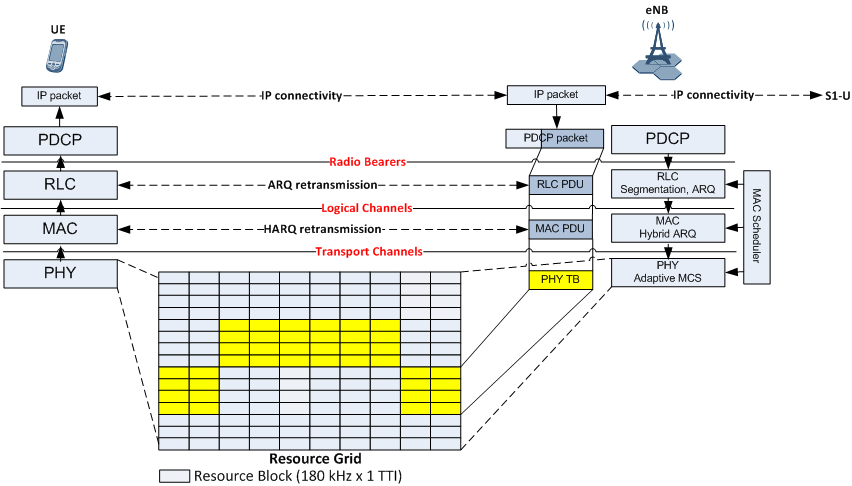}
	\caption{RAN network elements supporting mobile video delivery over 4G LTE.}
	\label{Fig_3}
	\end{figure}

\textbf{4G LTE Radio Access Network:} RAN comprises large number of eNBs establishing radio connections to user devices (UEs). Figure \ref{Fig_3} illustrates RAN user plane protocol stack for data delivery between eNB and UE. IP data flow towards UE passes through the Packet Data Conversion Protocol (PDCP) for header compression and ciphering, and Radio Link Control (RLC) protocol for segmentation/concatenation into suitably sized RLC packets that match the MAC frame size. If used in acknowledged automatic repeat request (ARQ) mode, RLC handles error-free and in-sequence RLC packet delivery between the eNB and UE. MAC layer introduces hybrid ARQ (HARQ) protection where, if MAC frame is not received correctly, up to three additional incremental redundancy MAC frames are transmitted. Finally, each MAC frame is allocated a single PHY transport block (PHY TB) placed into an OFDM-based time-frequency resource grid available to the eNB \cite{DPBook}.

PHY TBs of concurrent IP data flows are scheduled onto PHY resource blocks (PHY RBs) within every transmission time interval (TTI) of $1$ms duration. A single PHY RB is a resource allocation unit of $1$ TTI time duration and 12 OFDM carriers (180 kHz). The total number of PHY RBs per TTI depends on the bandwidth allocated to eNB (e.g., $50$ PHY RBs for $10$ MHz downlink channel). The information carrying capacity of PHY TBs depends on the number of PHY RBs allocated to the UE, adaptive modulation and coding scheme (MCS) applied and the multiple antenna (MIMO) mode used. Note that the user plane downlink and uplink data flows are multiplexed as part of the hierarchy of logical, transport and physical channels defined at LTE MAC and PHY layer. For example, at the lowermost layer, user data is carried by the physical downlink/uplink shared channel (PDSCH/PUSCH), along with a number of other physical control channels and channel reference signals. For a detailed overview of LTE RAN interface, we refer to more detailed exposition in \cite{DPBook}.

\textbf{Radio Access Network Support for Video Streaming/Downloading:} eNB/UE interface represents the critical link in end-to-end video delivery from content servers to mobile devices, both in terms of channel capacity and variability. Optimization of resource allocation is a key factor for efficient usage of available radio spectrum. In LTE, MAC scheduler is responsible for allocation of PHY RBs to active UEs based on feedback on their channel quality indicators (CQI), and is usually based on proportional-fair (PF) scheduling \cite{DPBook,LTESch}. However, standard MAC schedulers are typically oblivious to video traffic. As a tempting idea, large number of studies explored cross-layer optimized design with MAC schedulers directly using perceived video-quality information \citep{QoESch1, QoESch2, QoESch3, QoESch4}. Although such a schemes offer performance gains, due to complexity of cross-layer implementation, they are rarely applied in practical systems.     

\textbf{Radio Access Network Support for Video Multicasting/Broadcasting:} Besides CN upgrade, eMBMS requires additional radio resources for eMBMS service. Thus a new set of logical, transport and physical channels have to be configured at the RAN interface. For example, physical multicast channel (PMCH) is introduced to carry user plane eMBMS data to eMBMS users in the cell. Resource allocation, coding and modulation configuration at PMCH can be further optimized for efficient LTE video broadcasting \citep{OpteMBMS1, OpteMBMS2, OpteMBMS3}, although these prospects are rarely applied at mobile operators.  

\section{Random Linear Network Coding for Mobile Video Delivery \label{sec4}}

In this section, we bring together the material presented in previous two sections and discuss integration of RLNC sublayer across different layers of 4G LTE mobile video delivery environment.

\subsection{RLNC: Where Should It Be?}
\begin{figure}[ht]
	\centering
	\includegraphics[width=15.0cm]{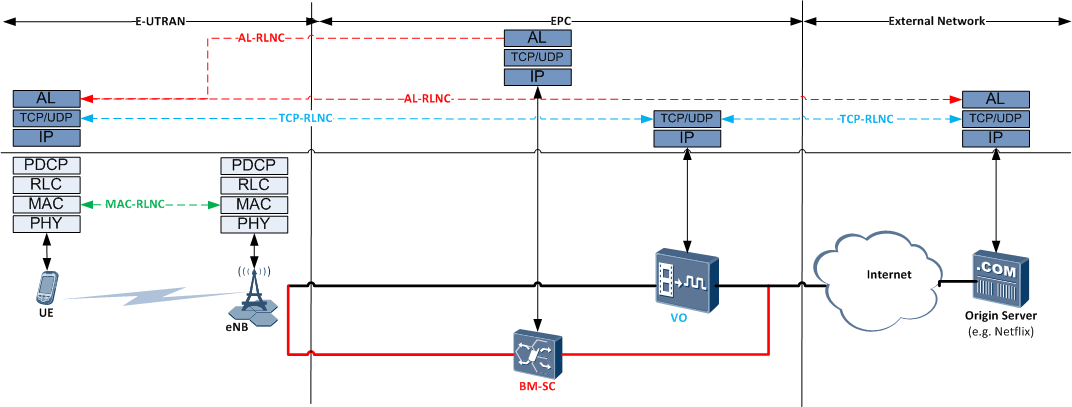}
	\caption{RLNC sublayer position options for mobile video delivery.}
	\label{Fig_4}
	\end{figure}

RLNC represents a flexible packet-level erasure coding sublayer that can be easily integrated at different positions within the protocol stack. RLNC flexibility rests on flexible definition of input source blocks, source block length $K$, source/encoded packet length $L$, coding coefficient field size $q$, number of encoded packets $N$ to be produced, and other RLNC properties such as sparsity, tunability and UEP. Furthermore, simple analytical expressions and bounds for packet decoding probabilities, outage probabilities and average decoding delays provide for optimized RLNC design in different scenarios, as detailed in Section \ref{sec2}. In the following, we discuss the suitable position for RLNC sublayer in the context of mobile video delivery. Possible opportunities for RLNC sublayer placement are divided into: (1) End-to-end solutions for RLNC residing either at application or transport layer; and (2) RLNC solutions residing within RAN protocol stack, as illustrated in Figure \ref{Fig_4}.

 \noindent  

\subsubsection{End-to-end Solutions for RLNC}

\textbf{Application Layer RLNC:} The simplest and easiest approach to integrate RLNC sublayer into mobile video delivery is to perform application-layer RLNC (AL-RLNC). In particular, for multicast/broadcast scenario (eMBMS), the required infrastructure is already in place, as eMBMS provides native support for AL-FEC \cite{eMBMS}. However, in this case, RLNC provides similar performance as compared to already proposed AL-FEC solutions such as Raptor codes \cite{RaptorQ} or LDPC triangle/staircase codes \cite{LDPCRFC}, providing limited benefit since no re-encoding at intermediate nodes is typically assumed in multicast/broadcast setup. Nevertheless, a large number of academic studies considered application and optimization of RLNC for mobile multicast video delivery demonstrating various benefits in different  scenarios \citep{NCMM, ARNC, MMTA, RLNCBCST}. Besides notable benefits of AL-RLNC in terms of flexibility and ease of implementation, there are several drawbacks of AL-RLNC worth emphasizing. AL-RLNC introduces redundancy either at the content server or BM-SC (eMBMS) thus adding significant communication overhead across the entire end-to-end IP multicast session, including typically reliable and over-provisioned core network optical links. In addition, AL-RLNC is transparent to lower layers and, in practice, it is hard to provide cross-layer optimized solution, e.g., at unreliable eNB/UE interface, that takes into account AL-RLNC. For unicast video streaming services, AL-RLNC is usually not considered as the most popular ABR/PD streaming solutions rely on HTTP via TCP, which already provides reliable packet delivery. However, this might change as more and more ABR traffic moves to Quick UDP Internet Connections (QUIC) protocol which relies on UDP \cite{QUIC}. However, note that most of the current unicast video traffic is encrypted, which might introduce practical limitations in applying RLNC at all layers below the application layer.  

\textbf{Transport Layer RLNC:} Another possibility for end-to-end RLNC is integration of RLNC sublayer in transport layer protocols. In the case of TCP, a groundbreaking idea has been investigated in which, instead of source packets, TCP transmits coded packets, and where instead of acknowledgement of individual source packets, the TCP receiver acknowledges received ``degrees of freedom'', i.e., the rank of the decoding matrix currently available at the receiver \cite{RLNCfeedback,CodedTCP}. Clearly, coded TCP is suitable solution for unicast ABR/PD video services which rely on TCP, thus any throughput improvements of coded TCP over traditional TCP would reflect on OTT video traffic. In addition, with VO platform in-the-middle (Section  \ref{sec3.2}), it is possible to split TCP connection between the ABR/PD streaming client and the content server, thus independently optimizing each of the two resulting TCP connections: the one between the UE and the VO platform, and the one between the VO platform and the content server. As a potential obstacle to coded TCP, its impact on TCP congestion control has yet to be better understood \cite{RLNCTCPReno}. 

\subsubsection{RLNC Solutions in Radio Access Network}

As an alternative to integration at the higher layers, RLNC sublayer could be integrated where the network reliability is critical: within RAN protocols. This solution could be equally applied for unicast and multicast/broadcast mobile video delivery, as it is triggered over the packets delivered between eNB and UE(s) either via unicast (PDSCH) or multicast (MTCH) transport channels.

\begin{figure}[H]
	\centering
	\includegraphics[width=13.0cm]{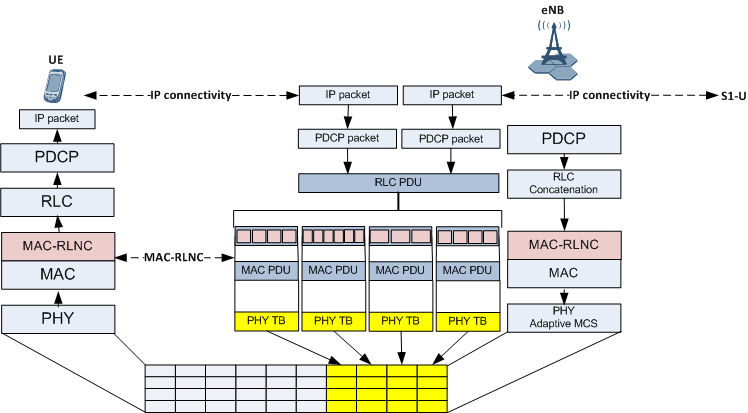}
	\caption{MAC-RLNC sublayer as part of LTE RAN protocol stack.}
	\label{Fig_5}
	\end{figure}

The solution for RLNC within the MAC layer of LTE RAN protocol stack has been investigated in detail in \cite{LTERLNC}. The proposed MAC-RLNC solution, integrated as the upper sublayer within MAC protocol, adopts the built-in flexibility of RLC protocol (segmentation and concatenation) to define the desired size of the source block that will be provided to the MAC-RLNC sublayer in the form of RLC packet data unit (PDU). From each received RLC PDU, MAC-RLNC produces a stream of carefully-sized fixed-length coded packet which are encapsulated in consecutive MAC frames and delivered via underlying PHY TB containers to the receiver side. The rationale behind MAC-RLNC was to apply rateless RLNC concept and convert RLC PDU transmission into a ``fluid'' delivery of coded packets flexible to fit PHY TB containers of different sizes. This way, MAC-RLNC would essentially replace MAC-based HARQ protocol, as instead of sending HARQ retransmissions, the transmitter simply continues to send a new set of coded packets, until receiver acknowledges it has received full-rank set and is able to recover the RLC PDU (see Figure \ref{Fig_5}). Performance of such a MAC-RLNC scheme has been investigated in terms of video delivery over LTE \cite{LTERLNCMM}. Integration of RLNC into RAN protocols opens a novel possibilities for joint optimization of RLNC and resource allocation for both unicast and multicast/broadcast services, the topic we present in more details in the following~subsection.   

 \noindent 

\subsection{RLNC and Resource Allocation}

Integration of RLNC sublayer within LTE protocol stack opens novel and interesting problems of interaction between RLNC and resource allocation that we describe next. 

We consider a general system model where a base station broadcasts an $\ell$-layer video stream encoded according to the SVC paradigm. We also say that layer $1$ is the basic and layers $2, \ldots, L$ are the enhancement layers. Assuming our system model adopts a NOW-RLNC implementation, any resource allocation strategy has to answer the following questions~\cite{TassiICC1,TassiTVT,TassiJSAC,TCL2016}:
\begin{itemize}
\item How many coded packet transmissions per-video layer are to be scheduled for transmission?
\item What MCSs are to be used for broadcasting each coded packet?
\end{itemize}
Obviously, choosing the correct number of coded packet transmission affects the user decoding probability. Likewise, the MCS selection has a direct impact on the total number of users capable of successfully decoding a given number of video layers.

Let $\mathcal{F}$ be our utility function, a General Resource Allocation Problem (GRAP) for network coded video application has the following structure:
\begin{align}
    \text{(GRAP)} &  \quad  \min_{\mathbf{m},\mathbf{N}} \,\, \mathcal{F} \label{GRAP.of}\\
    \text{subject to}                      &   \quad m_{\ell-1} < m_{\ell}, \quad\quad \text{$\ell = 2, \ldots, L$}\label{GRAP.c1}\\
 &   \quad U_\ell \geq \Hat{U}_\ell, \quad\quad\quad \text{$\ell = 1, \ldots, L$}\label{GRAP.c2}\\
                      &   \quad \mathcal{S}(\mathbf{N}) \leq \Hat{S} \label{GRAP.c3}\\
                      &   \quad \mathcal{T}(\mathbf{N}) \leq \Hat{T} \label{GRAP.c4}
\end{align}
where $m_\ell$ signifies the MCS to be used for transmitting coded packets associated with layer $\ell$. In addition, we define our optimization variables to be the vectors $\mathbf{m} = \{m_1,\ldots,m_L\}$ and $\mathbf{N} = \{N_1,\ldots,N_L\}$. Constraint~\eqref{GRAP.c1} ensures that coded packets associated with layer $\ell-1$ are being transmitted with a smaller MCS compared to that used for layer $\ell$ -- thus ensuring coded packets associated with layer $\ell-1$ are received with a PER $\epsilon$ that is smaller than or equal to that experienced in receiving coded packets associated with layer $\ell$. From~\eqref{eq2}, it follows that the probability of a user successfully recovering the first $\ell$ video layers can be expressed as $\prod_{i=1}^\ell P^i_d(N_i)$, where $N_i$ is the number of coded packet transmission associated with layer $i$. Let us signify with $U_\ell$, the number of users that can successfully decode the first $\ell$ video layers with a probability equal to or greater than $\Hat{p}_d$. Constraint~\eqref{GRAP.c2} ensures that $U_\ell$ is greater than or equal to a target number of users $\Hat{U}_\ell$. We observe that relation $\Hat{U}_{\ell-1} \geq \Hat{U}_\ell$ holds true for $\ell = 2, \ldots, L$ as it would be pointless to impose video layer $\ell$ to be decoded by a number of users larger than $\ell-1$. Constraint~\eqref{GRAP.c3} ensures that the total number of coded packets scheduled on each radio frame does not exceed a given threshold $\Hat{S}$. Finally, constraint~\eqref{GRAP.c4} ensure that each portion of video stream is transmitted by a time $\Hat{T}$.

Unfortunately, to the best of our knowledge, GRAP is an integer (potentially) non-linear optimization problem with no obvious analytical solutions. However, when $\mathcal{F}$ is equal to the total number of coded packet transmissions, authors'~\cite{TassiTVT,TassiJSAC} argued that it is possible to find a good-quality feasible solution to GRAP using a two-step heuristic operating as follows:
\begin{enumerate}
\item \emph{MCS Allocation}---The heuristic iterates over all the MCSs (starting from the highest), and for each video layer, it identifies the largest MCS such that constraints~\eqref{GRAP.c1} and~\eqref{GRAP.c2} are met. The output of this heuristic step is an instance of vector $\mathbf{m}$.
\item \emph{Code Packet Transmissions Optimisation}---On the basis of the instance of vector $\mathbf{m}$ determined in the previous step, the minimum value of $N_\ell$ is determined, for any $\ell = 1,\ldots,L$---thus an instance of vector $\mathbf{N}$ is found.
\end{enumerate}

It is easy to prove the aforementioned heuristic has a reduced computational complexity and when it identifies an instance of $\mathbf{m}$ and $\mathbf{N}$ they jointly constitute a feasible solution to GRAP. Since $\mathbf{m}$ and $\mathbf{N}$ are independently optimized, the heuristic solution is not always optimum. However, for realistic network and video service deployments, the heuristic provides good-quality solutions~\cite{TassiJSAC,TassiICC}.

It has also been observed that GRAP can be extended to system models where EW-RLNC implementations are in use~\cite{TassiJSAC}. In these cases, terms $N_1, \ldots, N_L$ represents the number of coded packet transmissions associated with expanding windows $\mathbf{w_1},\ldots,\mathbf{w_L}$ to be scheduled for transmission. Similarly, terms $m_1, \ldots, m_L$ identifies the MCSs to be used for transmissions associated with $\mathbf{w_1},\ldots,\mathbf{w_L}$, respectively. As observed in Section~\ref{sec.UEP-RLNC}, a user can decode the first $\ell$ video layers if it successfully recovers any of the expanding windows $\mathbf{w_\ell},\mathbf{w_\ell+1},\ldots,\mathbf{w_L}$. As such, in this case, the probability of a user successfully recovering the first $\ell$ video layers is $\bigvee_{i=\ell}^L P_{d,\mathbf{w_i}}(\mathbf{N})$, where $P_{d,\mathbf{w_i}}(\mathbf{N})$ is the probability of a user recovering $\mathbf{w_i}$ -- thus terms $\mathcal{U}_1, \ldots, \mathcal{U}_L$ have to be redefined accordingly. Despite its new setting, the aforementioned two-step heuristic can still be applied to find good-quality feasible solution to GRAP~\cite{TassiJSAC}.

\section{Random Linear Network Coding for 5G New Radio \label{sec5}}

Standardization of 5G NR technology is currently under way within 3GPP. In the moment of writing of this paper, the Phase 1 of the 5G NR standardization has just been completed \cite{5GNR}. In this section, we provide a short intro to 5G NR, followed by identification of several 5G NR standard features suitable for further study of RLNC integration and optimization of mobile video delivery in 5G. We conclude the section with recent related work on video delivery deployment studies in 5G.  

\subsection{Introduction to 5G NR}

\begin{figure}[H]
	\centering
	\includegraphics[width=9.0cm]{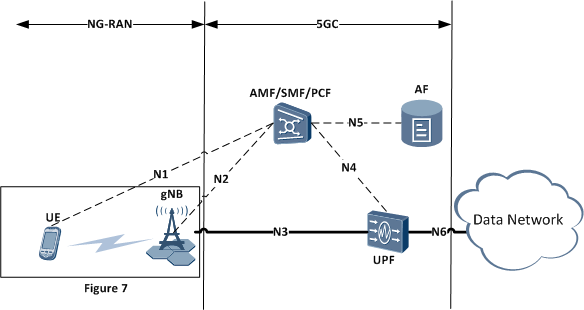}
	\caption{5G NR System Architecture.}
	\label{Fig_6}
	\end{figure}

\textbf{Overall 5G NR Architecture:}  The 5G system architecture consists of a 5G Access Network (AN), 5G Core (5GC) Network and the UE \cite{5GArchitecture}. The 5G AN comprises an NG-RAN and/or non-3GPP AN connecting to a 5G Core Network. NG-RAN focuses on the radio interface protocol architecture and contains NG-RAN nodes termed next-generation NodeB (gNB), providing NR user plane and control plane protocol terminations towards the UE. The gNBs are interconnected with each other and are also connected by means of the NG interfaces to the 5GC, most importantly to the AMF (Access and Mobility Management Function) and to the UPF (User Plane Function), as illustrated in Figure \ref{Fig_6} \cite{TS38300}. For the details on 5G NR architecture and other 5GC entities and interfaces, we refer the interested reader to \cite{5GArchitecture}.

 \noindent 

\begin{figure}[H]
	\centering
	\includegraphics[width=12.0cm]{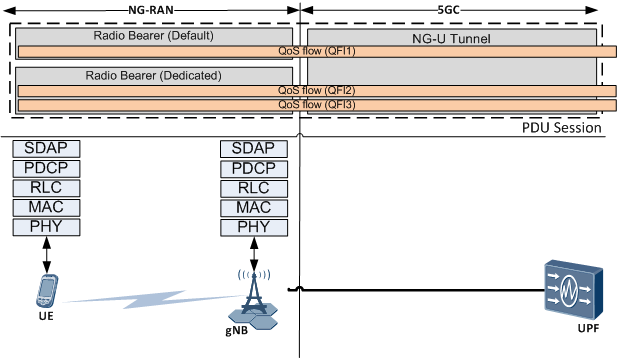}
	\caption{5G NG-RAN Protocol Stack and 5G QoS Architecture.}
	\label{Fig_7}
	\end{figure}

\textbf{5G NR Radio Protocol Architecture - User Plane:} Figure \ref{Fig_7} (left) illustrates the protocol stack for the gNB user plane that includes Service Data Adaptation Protocol (SDAP), PDCP, RLC, MAC and PHY sublayers. The new SDAP sublayer is introduced to 5G NR targeting specifically Quality of Service (QoS) control in NG-RAN. SDAP handles: i) the mapping of 5G QoS flows to data radio bearers (DRBs), and ii) marking of 5G QoS flow Identifier (QFI) to a QoS flow in both DL and UL packets. We will discuss in more detail 5G QoS architecture later in this subsection.

The main functions of PDCP sublayer remain similar to 4G LTE (e.g., header compression, ciphering and integrity protection), with the exception of one important new feature called PDCP PDU duplication. The PDCP duplication offers the possibility of sending the same PDCP PDUs twice: once on the original RLC entity and a second time on the additional RLC entity. The original and duplicate PDCP PDUs are transmitted on different carriers. The two different logical channels can either belong to the same MAC entity, e.g., via carrier aggregation (CA), or to different logical channels in the case of multi-connectivity, i.e., dual connectivity (DC) \cite{TS38300}. As we discuss later, combined with the PDCP duplication feature, PDCP could provide an ideal place for RLNC sublayer integration.

The RLC protocol preserved most of the LTE functionalities, including ARQ-based error correction, segmentation and reassembly of SDUs, etc. Similarly, the main services and functions of the MAC sublayer resemble those in LTE and include, e.g., error correction through HARQ, priority handling between UEs by means of dynamic scheduling, and priority handling between logical channels of one UE by means of logical channel prioritisation.

\textbf{QoS Architecture in 5G NR:}  Basic granularity for QoS control in LTE EPC/E-UTRAN is EPS bearer/E-UTRAN Radio Access Bearer (E-RAB). EPS bearer/E-RAB established when UE connects to a PDN is called the default bearer, while any additional bearer is referred to as a dedicated bearer. Each  bearer is characterized by the same packet forwarding treatment (e.g., scheduling policy, queue management policy, rate shaping policy, RLC configuration, etc.). A bearer is called guaranteed bit-rate (GBR) bearer if it is provided dedicated network resources, otherwise, it is called non-GBR bearer.

 \noindent 

In 5G, the concept of QoS flow is introduced which is considered the finest data flow granularity that receives the same forwarding treatment. Providing different QoS forwarding treatment requires separate 5G QoS flows. Multiple QoS flows are part of a packet data unit (PDU) session: an association between the UE and a PDN \cite{5GArchitecture}.  For each UE, the 5GC may establish one or more PDU sessions carrying QoS flows of different QoS level. QoS flow ID (QFI) identifies a QoS flow within a PDU session and indicates a specific QoS forwarding behavior (e.g. packet loss rate, packet delay budget). At the NG-RAN, the QoS flows are mapped to one or more data radio bearers (DRBs). NG-RAN and 5GC jointly ensure QoS by mapping packets to appropriate QoS flows and DRBs by a 2-step mapping process, where an IP-flow is mapped by 5GC to QoS flows while NG-RAN (SDAP protocol) maps QoS flows to DRBs. Within each PDU session, NG-RAN (SDAP) decides how to map QoS flows to DRBs. The QoS architecture in 5G is illustrated in Figure \ref{Fig_7}.

\subsection{Opportunities and Challenges for Mobile Video Delivery in 5G}

\textbf{PDCP Coded Duplication in 5G NR:} Due to PDCP duplication feature, PDCP protocol becomes an interesting option for the RLNC sublayer within the NG-RAN user plane protocol stack. Empowering PDCP layer with RLNC sublayer could provide a possibility for ``coded duplication'' where, instead of a simple duplication of PDCP PDUs, one could transmit two different sets of RLNC coded packets created from the appropriately defined source block. Next, we briefly describe the concept of coded duplication but leave the more detailed investigation for our future work.

\begin{figure}[H]
	\centering
	\includegraphics[width=12.0cm]{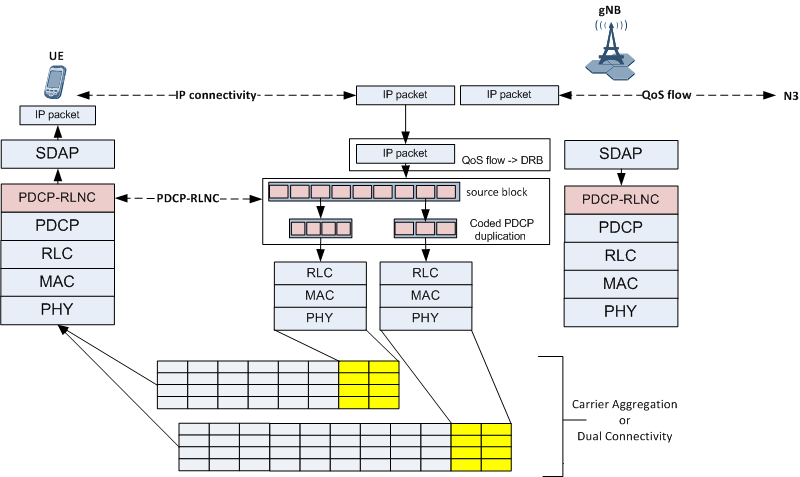}
	\caption{The concept of Coded PDCP Duplication in 5G NR.}
	\label{Fig_8}
	\end{figure}

Figure \ref{Fig_8} illustrates the concept of PDCP Coded Duplication. PDCP protocol receives PDCP service data units (SDUs), i.e., IP packets, and organizes one or more PDCP SDUs into a source block of length $K$ source packets. The source block is processed by RLNC sublayer to produce $N$ coded packets encapsulated into a sequence of coded PDCP PDUs. The set of PDCP PDUs carrying the coded packets of the same source block are enumerated using the same PDCP sequence number and transmitted using PDCP duplication feature via separate protocol legs. The coded duplicated content  propagates down the two independent legs, undergoing possibly different segmentation at the two RLC entities followed by different MAC frame delivery success. Note that due to RLNC coding at PDPC layer, RLC layer ARQ and MAC layer HARQ could be altogether avoided in coded PDCP duplication. 

At the receiving side, instead of removal of PDCP duplicates, the receiving PDCP entity would collect the coded content from all incoming PDCP PDUs tagged with the same sequence number and received through both legs. The coded packets are extracted from PDCP PDU until the content of the source block is reconstructed, as illustrated in Figure \ref{Fig_8}.  Finally, note that coded PDCP duplication operates across two parallel packet erasure channels, which offers different possibilities for matching layered UEP RLNC schemes and parallel erasure channels, empowered with appropriate resource allocation of coded packets to coded duplicated PDCP PDUs. 

\textbf{QoS Control for 5G Mobile Video Delivery:}  The 5G QoS architecture described in the previous subsection will provide advanced mechanisms for QoS control in mobile video delivery. 5GC will classify QoS flows based on their QFI and provide per flow forwarding treatment in terms of e.g. packet loss rates and packet delay, thus providing mobile video flows with desired delivery parameters. 

One of the key elements of 5G QoS architecture is a novel SDAP protocol defined at NG-RAN user plane interface. SDAP is additional resource allocation entity in 5G NR that will complement MAC scheduler. While the ``lower layer'' MAC scheduler aims to dynamically schedule resource blocks to different UEs in order to maintain guaranteed or best possible DRB parameters, the ``upper layer'' SDAP scheduler aims to assign QoS flows to different DRBs in order to satisfy their QFI-defined parameters. Note that different DRBs defined at the interface towards the UE may be configured using different NG-RAN protocol configurations. In this sense, coded PDCP duplication described above may provide a flexible approach to trade off reliability and latency and fine tune QFI-defined requirements in terms of packet loss rates and packet delay for a given DRB. 

\subsection{Related Studies on Mobile Video Delivery in 5G NR}

Without the goal of being exhaustive, we finalize this review paper with an overview of related work in the domain of mobile video delivery in 5G. Along with our discussion above, these papers could help the reader to identify possible research avenues in the domain of 5G mobile video delivery.

Starting with the  general papers on 5G architecture and services, the METIS project vision presents eMBB service that will support future UHD, multi-view or 360-degree video services \cite{METIS}. For the core network, mobile edge computing (MEC) and edge content caching are identified as promising NG-RAN architectures that will greatly improve massive unicast OTT video delivery services such as ABR.  In \cite{VideoMEC}, the combination of network function virtualization (NFV) and MEC is explored for deployment of context-aware virtualized adaptive prefetching agents at the mobile edge that will provide QoS-guaranteed UHD video services.   The work in \cite{5GQoS} focuses on QoS provisioning in 5G mobile networks with emphasis on mobile video delivery services. In ultra-dense 5G network scenario, the work presented in \cite{Femtocaching} advocates local caching of video in small cells in combination with device-to-device (D2D) communication, while in \cite{IEEENetwork2017}, the authors consider network-aware ABR video content caching at the network edge combined with end-to-end video streaming resource allocation optimization. Edge caching is jointly optimized with multicast transmissions in what authors term as multicast-aware edge caching for 5G video delivery of popular content in \cite{MACaching}. We also point to the multi-server video streaming architecture for optimized ABR video delivery in 5G, relying on a combination of cloud RAN (C-RAN) and  MEC concepts, as detailed in \cite{HAL}.  

In RAN domain, we emphasize several studies related to the work presented here. The work in \cite{NCmmWave} presents a study on high-quality high-throughput video streaming with enhanced reliability, which is achieved using a combination of end-to-end RLNC and 5G multi-connectivity via legacy LTE and 5G mmWave radio links.   The combination of mmWave radio links and video caching at network edge is investigated as suitable solution for 5G mobile video in \cite{5GShen}. Our discussion on coded PDCP duplication draws inspiration from the recent work on optimized interface diversity for 5G ultra-reliable and low-latency services (URLLC) \cite{URLLC}. Both studies point out to the potential of using multi-connectivity (e.g., via carrier aggregation or dual-connectivity) for reliable and low-delay video delivery in 5G. Finally, we also emphasize the role of resource management in NG-RAN, as recently explored in 5G mobile vehicular video delivery in \cite{PIMRC}. Resource management in NR-RAN will require novel ideas for optimized cross-layer video optimization exploiting both SDAP-based allocation of QoS flows to DRBs and MAC-based dynamic scheduling of radio resources to UE terminals.

\section{Conclusions \label{sec6}}

The goal of this paper was to present a detailed study of two interrelated topics, RLNC for packet erasure protection and mobile video delivery in mobile cellular systems. We introduced RLNC using module-based approach, motivated by the quest to identify both the need and the suitable location for RLNC sublayer in video delivery solutions for 4G/5G mobile cellular networks. Evolving from RLNC sublayer integration in 4G LTE, the paper culminates with investigation of 5G NR architecture and possible RLNC integration therein for future 5G optimized mobile video delivery. Across the paper, we used the opportunity to provide detailed review and point towards relevant publications of all the fundamental concepts related to mobile video delivery in 4G/5G networks.  
\vspace{6pt}

\acknowledgments{All sources of funding of the study should be disclosed. Please clearly indicate grants that you have received in support of your research work. Clearly state if you received funds for covering the costs to publish in open access.}

\authorcontributions{For research articles with several authors, a short paragraph specifying their individual contributions must be provided. The following statements should be used ``X.X. and Y.Y. conceived and designed the experiments; X.X. performed the experiments; X.X. and Y.Y. analyzed the data; W.W. contributed reagents/materials/analysis tools; Y.Y. wrote the paper.'' Authorship must be limited to those who have contributed substantially to the work reported.} 

\conflictsofinterest{The authors declare no conflict of interest.}  




\reftitle{References}


\begin{thebibliography}{999}

\bibitem{VNI}
Cisco Visual Networking Index. Available online: \url{https://www.cisco.com/c/en/us/solutions/service-provider/
/visual-networking-index-vni} (accessed on Day Month Year). 

\bibitem{UHD}
\textls[-5]{Ye, Y.; Andrivon, P. The scalable extensions of HEVC for ultra-high-definition video delivery. \emph{IEEE~MultiMedia}} \textbf{2014}, \emph{21}, 58--64.

\bibitem{5GVR}
Bastug, E.; Bennis, M.; Médard, M.; Debbah, M. Toward interconnected virtual reality: Opportunities, challenges, and enablers. \emph{IEEE Commun. Mag.} \textbf{2017}, \emph{55}, 110--117.

\bibitem{BCASTMM}
De la Fuente, A.; Leal, R.P.; Armada, A.G. New technologies and trends for next generation mobile broadcasting services. \emph{IEEE Commun. Mag.} \textbf{2016}, \emph{54}, 217--223.

\bibitem{P2PMM}
Xu, C.; Jia, S.; Zhong, L.;  Muntean, G.M. Socially aware mobile peer-to-peer communications for community multimedia streaming services. \emph{IEEE Commun. Mag.} \textbf{2015}, \emph{53}, 150--156.

\bibitem{5GNR}
3GPP 5G NR Rel. 15 Specification Series. Available online: \url{http://www.3gorg/DynaReport/38-series.htm} (accessed on Day Month Year) 

\bibitem{5GRAN}
Marsch, P.; Da Silva, I.; Bulakci, O.; Tesanovic, M.; El Ayoubi, S.E.; Rosowski, T.; Kaloxylos, A.;  Boldi, M. 5G radio access network architecture: Design guidelines and key considerations. \emph{IEEE~Commun. Mag.}  \textbf{2016}, \emph{54}, 24--32.

\bibitem{5GCN}
Rost, P.; Banchs, A.; Berberana, I.; Breitbach, M.; Doll, M.; Droste, H.; Mannweiler, C.; Puente, M.A.; Samdanis, K. and Sayadi, B. Mobile network architecture evolution toward 5G. \emph{IEEE Commun. Mag.} \textbf{2016}, \emph{54}, 84--91.

\bibitem{eMBMS}
Lecompte, D.; Gabin, F. Evolved multimedia broadcast/multicast service (eMBMS) in LTE-advanced: overview and Rel-11 enhancements. \emph{IEEE Commun. Mag.} \textbf{2012}, \emph{50}, 68--74.

\bibitem{QoEVoLTE}
Oyman, O.; Foerster, J.; Tcha, Y.J.; Lee, S.C. Toward enhanced mobile video services over WiMAX and LTE. \emph{IEEE Commun. Mag.}  \textbf{2011}, \emph{48}, doi:10.1109/MCOM.2010.5534589.

\bibitem{QoESch1}
Luo, H.; Ci, S.; Wu, D.; Wu, J.; Tang, H. Quality-driven cross-layer optimized video delivery over LTE. \emph{IEEE~Commun. Mag.} \textbf{2010},  \emph{48}, doi:10.1109/MCOM.2010.5402671.

\bibitem{EdgeCache}
Wang, X.; Chen, M.; Taleb, T.; Ksentini, A.; Leung, V. Cache in the air: exploiting content caching and delivery techniques for 5G systems. \emph{IEEE Commun. Mag.} \textbf{2014}, \emph{52}, 131--139.

\bibitem{LontheEdge}
Bastug, E.; Bennis, M.; Debbah, M. Living on the edge: The role of proactive caching in 5G wireless networks. \emph{IEEE Commun. Mag.} \textbf{2014}, \emph{52}, 82--89.

\bibitem{MEC}
Tran, T.X.; Hajisami, A.; Pandey, P.; Pompili, D. Collaborative mobile edge computing in 5G networks: New paradigms, scenarios, and challenges. \emph{IEEE Commun. Mag.} \textbf{2017}, \emph{55}, 54--61.

\bibitem{HetAl2006}
Ho, T.; M\`{e}dard, M.; Koetter, R.; Karger, D.R.; Effros, M.; Shi, J.; Leong, B. A random linear network coding approach to multicast. \emph{IEEE Trans. Inf. Theory} \textbf{2006}, \emph{52}, 4413--4430.

\bibitem{FBW2006}
Fragouli, C.; Le Boudec, J.Y.; Widmer, J. Network coding: An instant primer. \emph{ACM SIGCOMM Comput. Commun. Rev.} \textbf{2006}, \emph{36}, 63--68.

\bibitem{CW2007}
Chou, P.A.; Wu, Y. Network coding for the internet and wireless networks. \emph{IEEE Signal Process. Mag.} \textbf{2007}, \emph{24}, 77--85.

\bibitem{FSbook2007}
Fragouli, C.; Soljanin, E. Network coding fundamentals. \emph{Found. Trends  Netw.} \textbf{2007}, \emph{2}, 1--133.

\bibitem{TCL2016}
Tassi, A.; Chatzigeorgiou, I.; Lucani, D.E. Analysis and optimization of sparse random linear network coding for reliable multicast services. \emph{IEEE Trans. Commun.} \textbf{2016}, \emph{64}, 285--299.

\bibitem{BJT2018}
Brown, S.; Johnson, O.; Tassi, A. Reliability of Broadcast Communications Under Sparse Random Linear Network Coding.  \emph{IEEE Trans. Veh. Technol.} \textbf{2018}, to appear.

\bibitem{Feizi2014}
Feizi, S.; Lucani, D.E.; Sørensen, C.W.; Makhdoumi, A.; M\`{e}dard, M. Tunable sparse network coding for multicast networks. In Proceedings of the  2014 International Symposium on Network Coding (NetCod) 2014, Aalborg, Denmark , 27--28 June 2014.

\bibitem{VS2012}
Vukobratovic, D.; Stankovic, V. Unequal error protection random linear coding strategies for erasure channels. \emph{IEEE Trans. Commun.} \textbf{2012}, \emph{60}, 1243--1252.

\bibitem{svccomm}
Schierl, T.; Stockhammer, T.; Wiegand, T. Mobile video transmission using scalable video coding. \emph{IEEE Trans. Circ. Syst. Video Technol.} \textbf{2007}, \emph{17}, 1204--1217.

\bibitem{PD}
Ma, K.J.; Bartos, R.; Bhatia, S.; Nair, R. Mobile video delivery with HTTP. \emph{IEEE Commun. Mag.}  \textbf{2011}, \emph{49},  doi:10.1109/MCOM.2011.5741161.

\bibitem{ABR}
Oyman, O.; Singh, S. Quality of experience for HTTP adaptive streaming services. \emph{IEEE Commun. Mag.} \textbf{2012},  \emph{50}, doi:10.1109/MCOM.2012.6178830 .

\bibitem{C2000}
Cooper, C. On the distribution of rank of a random matrix over a finite field. \emph{Random Struct. Algorithms} \textbf{2000}, \emph{17}, 197--212.

\bibitem{TC2011}
Trullols-Cruces, O.; Barcelo-Ordinas, J.M.; Fiore, M. Exact decoding probability under random linear network coding. \emph{IEEE Commun. Lett.} \textbf{2011}, \emph{15}, 67--69.

\bibitem{Nistor2011}
Nistor, M.; Lucani, D.E.; Vinhoza, T.T.; Costa, R.A.; Barros, J. On the delay distribution of random linear network coding. \emph{IEEE J. Sel. Areas  Commun.} \textbf{2011}, \emph{29}, 1084--1093.

\bibitem{CT2017}
Chatzigeorgiou, I.; Tassi, A. Decoding delay performance of random linear network coding for broadcast. \emph{IEEE Trans. Veh. Technol.} \textbf{2017}, \emph{66}, 7050--7060.

\bibitem{LPC2010}
Liva, G.; Paolini, E.; Chiani, M. Performance versus overhead for fountain codes over $\mathbb{F}_q$. \emph{\mbox{IEEE Commun. Lett.}} \textbf{2010}, \emph{14}, 178--180.

\bibitem{LT2002}
Luby, M. LT codes. In Proceedings of the 43rd Annual Symposium on Foundations of Computer Science (FOCS 2002), Vancouver, BC, Canada, 16--19 November 2002; pp. 271--280.

\bibitem{Raptor2006}
Shokrollahi, A. Raptor codes. \emph{IEEE Trans. Inf. Theory} \textbf{2006}, \emph{52}, 2551--2567.

\bibitem{PF2005}
Pakzad, P.; Fragouli, C.; Shokrollahi, A. Coding schemes for line networks. In Proceedings of the  International Symposium on Information Theory, 2005 (ISIT 2005),  Adelaide,  Australia,  4--9 September 2005;    pp.~1853--1857.

\bibitem{LMK2006}
Lun, D.S.; M\`{e}dard, M.; Koetter, R. Network coding for efficient wireless unicast. In Proceedings of the  2006 International Zurich Seminar on Communications, Zurich, Switzerland, 22--24 February 2006; pp. 74--77.

\bibitem{SMR2011}
Seferoglu, H.; Markopoulou, A.; Ramakrishnan, K.K. $I^{2}NC$: Intra-and inter-session network coding for unicast flows in wireless networks. In Proceedings of the IEEE INFOCOM 2011, Shanghai, China, \mbox{10--15 April 2011}; pp. 1035--1043.

\bibitem{h265hevc}
Sullivan, G.J.; Ohm, J.; Han, W.J.; Wiegand, T. Overview of the high efficiency video coding (HEVC) standard. \emph{IEEE Trans. Circ. Syst. Video Technol.} \textbf{2012}, \emph{22}, 1649--1668.

\bibitem{h264avc}
Wiegand, T.; Sullivan, G.J.; Bjontegaard, G.; Luthra, A. Overview of the H. 264/AVC video coding standard. \emph{IEEE Trans. Circ. Syst. Video Technol.} \textbf{2007}, \emph{13}, 560--576.

\bibitem{hevcExt}
Sullivan, G.J.; Boyce, J.M.; Chen, Y.; Ohm, J.R.; Segall, C.A.; Vetro, A. Standardized extensions of high efficiency video coding (HEVC). \emph{IEEE J. Sel. Top. Signal Process.} \textbf{2013}, \emph{7}, 1001--1016.

\bibitem{Nazir2015}
Nazir, S.; Vukobratović, D.; Stanković, V.; Andonović, I.; Nybom, K.; Groenroos, S.  Unequal error protection for data partitioned H. 264/AVC video broadcasting. \emph{Multimed. Tools Appl.} \textbf{2015}, \emph{74}, 5787--5809.

\bibitem{CMGG2013}
Calabuig, J.; Monserrat, J.F.; Goz\`{a}lvez, D.; G\`{o}mez-Barquero, D. AL-FEC for streaming services in lte e-MBMS. \emph{EURASIP J. Wirel. Commun. Netw.} \textbf{2013}, \emph{2013}, 73.

\bibitem{LTE-B-WP}
LTE Broadcast---Lessons Learned from Trials and Early Deployments, LTE Alliance, Whitepaper, 2016.  Available online: \url{http://www.expway.com/lte-broadcast-lessons-learned-from-trials-and-early-deployments/} (accessed on Day Month Year).  

\bibitem{TS23.002}
3GPP TS 23.002, 14.1.0 Network Architecture. 

\bibitem{DPBook}
Dahlman, E.; Parkvall, S.; Skold, J. \emph{4G: LTE/LTE-Advanced for Mobile Broadband};  Academic Press: Cambridge, MA, USA, 2013. 

\bibitem{PV2012}
Pu, W.; Zou, Z.; Chen, C.W. Video adaptation proxy for wireless dynamic adaptive streaming over HTTP.  In Proceedings of the 2012 19th International Packet Video Workshop (PV), Munich, Germany, 10--11 May 2012;  pp. 65--70.

\bibitem{E2015}
El Essaili, A.; Schroeder, D.; Steinbach, E.; Staehle, D.; Shehada, M. QoE-based traffic and resource management for adaptive HTTP video delivery in LTE. \emph{IEEE Trans. Circ. Syst. Video Technol.} \textbf{2015}, \emph{25}, 988--1001.

\bibitem{Caching}
Zhang, W.; Wen, Y.; Chen, Z.; Khisti, A. QoE-driven cache management for HTTP adaptive bit rate streaming over wireless networks. \emph{IEEE Trans. Multimed.} \textbf{2013}, \emph{15}, 1431--1445.

\bibitem{Gorilla}
Erman, J.; Gerber, A.; Ramadrishnan, K.K.; Sen, S.; Spatscheck, O. Over the top video: The gorilla in cellular networks.  In Proceedings of the ACM SIGCOMM 2011, Berlin, Germany, 2--4 November  2011; pp. 127--136.

\bibitem{Medieval}
Amram, N.; Fu, B.; Kunzmann, G.; Melia, T.; Munaretto, D.; Randriamasy, S.; Sayadi, B.; Widmer, J.; Zorzi, M. QoE-based transport optimization for video delivery over next generation cellular networks. In Proceedings of the 2011 IEEE Symposium on Computers and Communications (ISCC), Kerkyra, Greece, 28 June--1 July 2011; pp. 19--24.

\bibitem{ALFECeMBMS}
Bouras, C.; Kanakis, N.; Kokkinos, V.; Papazois, A. AL-FEC for streaming services over LTE systems. In Proceedings of the 2011 14th International Symposium on Wireless Personal Multimedia Communications (WPMC),  Brest, France, 3--7 October 2011.

\bibitem{LTESch}
Capozzi, F.; Piro, G.; Grieco, L.A.; Boggia, G.; Camarda, P.  Downlink packet scheduling in LTE cellular networks: Key design issues and a survey. \emph{IEEE Commun. Surv. Tutor.} \textbf{2013}, \emph{15}, 678--700.

\bibitem{QoESch2}
Piro, G.; Grieco, L.A.; Boggia, G.; Fortuna, R.; Camarda, P. Two-level downlink scheduling for real-time multimedia services in LTE networks. \emph{IEEE Trans. Multimed.} \textbf{2011}, \emph{13}, 1052--1065.

\bibitem{QoESch3}
Lai, W. K.; Tang, C.L. QoS-aware downlink packet scheduling for LTE networks. \emph{Comput. Netw.}  \textbf{2013}, \emph{57}, 1689--1698.

\bibitem{QoESch4}
Su, G. M.; Su, X.; Bai, Y.; Wang, M.; Vasilakos, A. V.; Wang, H.  QoE in video streaming over wireless networks: perspectives and research challenges. \emph{Wirel. Netw.} \textbf{2016}, \emph{22}, 1571--1593.

\bibitem{OpteMBMS1}
Araniti, G.; Condoluci, M.; Militano, L.; Iera, A. Adaptive resource allocation to multicast services in LTE systems. \emph{IEEE Trans. Broadcast.} \textbf{2013}, \emph{59}, 658--664.

\bibitem{OpteMBMS2}
Chen, J.; Chiang, M.; Erman, J.; Li, G.; Ramakrishnan, K.K.; Sinha, R.K. Fair and optimal resource allocation for LTE multicast (eMBMS): Group partitioning and dynamics. In Proceedings of the 2015 IEEE Conference on Computer Communications (INFOCOM), Hong Kong, China,  26 April--1 May 2015; pp. 1266--1274.

\bibitem{OpteMBMS3}
Lau, C.P.; Alabbasi, A.; Shihada, B. An efficient live TV scheduling system for 4G LTE broadcast. \emph{IEEE Syst. J.} \textbf{2017}, \emph{11}, 2737--2748.

\bibitem{RaptorQ}
Bouras, C.; Kanakis, N.; Kokkinos, V.; Papazois, A. Embracing RaptorQ FEC in 3GPP multicast services. \emph{Wirel. Netw.} \textbf{2013}, \emph{19}, 1023--1035.

\bibitem{LDPCRFC}
IETF RFC 5170 Low Density Parity Check (LDPC) Staircase and Triangle Forward Error Correction (FEC) Schemes. Available online: \url{https://www.rfc-editor.org/rfc/rfc5170.txt}  (accessed on Day Month Year).  

\bibitem{NCMM}
Magli, E.; Wang, M.; Frossard, P.; Markopoulou, A. Network coding meets multimedia: A review. \emph{IEEE~Trans.~Multimed.} \textbf{2013}, \emph{15}, 1195--1212.

\bibitem{ARNC}
Li, B.; Li, H.; Zhang, R. Adaptive random network coding for multicasting hard-deadline-constrained prioritized data. \emph{IEEE Trans. Veh. Technol.} \textbf{2016}, \emph{65}, 8739--8744.

\bibitem{MMTA}
Shin, H.; Park, J.S. Optimizing random network coding for multimedia content distribution over smartphones. \emph{Multimed. Tools   Appl.} \textbf{2017}, \emph{76}, 19379--19395.

\bibitem{RLNCBCST}
Esmaeilzadeh, M.; Sadeghi, P.; Aboutorab, N. Random linear network coding for wireless layered video broadcast: General design methods for adaptive feedback-free transmission. \emph{IEEE Trans. Commun.} \textbf{2017}, \emph{65}, 790--805.

\bibitem{QUIC}
Carlucci, G.; De Cicco, L.; Mascolo, S. HTTP over UDP: An Experimental Investigation of QUIC. In Proceedings of the 30th Annual ACM Symposium on Applied Computing,  Salamanca, Spain, \mbox{13--17 April  2015}; pp. 609--614.

\bibitem{RLNCfeedback}
Fragouli, C.; Lun, D.; M\`{e}dard, M.; Pakzad, P. On feedback for network coding. In Proceedings of the 41st Annual Conference on Information Sciences and Systems,   Baltimore, MD, USA,  14--16 March 2007; pp.~248--252. 

\bibitem{CodedTCP}
Sundararajan, J.K.; Shah, D.; M\`{e}dard, M.; Mitzenmacher, M.; Barros, J. Network coding meets TCP.  In Proceedings of the  IEEE INFOCOM 2009,  Rio de Janeiro, Brazil, 19--25 April 2009; pp.  280--288.

\bibitem{RLNCTCPReno}
Medina Ruiz, H.; Kieffer, M.; Pesquet-Popescu, B.; Medina Ruiz, H.; Kieffer, M.; Pesquet-Popescu, B. TCP and Network Coding: Equilibriu and Dynamic Properties. \emph{IEEE/ACM Trans. Netw.} \textbf{2016}, \emph{24}, 1935--1947.

\bibitem{LTERLNC}
Khirallah, C.; Vukobratovic, D.; Thompson, J. Performance analysis and energy efficiency of random network coding in LTE-advanced. \emph{IEEE Trans. Wirel. Commun.} \textbf{2012}, \emph{11}, 4275--4285.

\bibitem{LTERLNCMM}
Vukobratovic, D.; Khirallah, C.; Stankovic, V.; Thompson, J.S. Random network coding for multimedia delivery services in LTE/LTE-Advanced. \emph{IEEE Trans. Multimed.} \textbf{2014}, \emph{16}, 277--282.

\bibitem{TassiTVT}
Tassi, A.; Khirallah, C.; Vukobratovic, D.; Chiti, F.; Thompson, J.; Fantacci, R. Resource Allocation Strategies for Network-Coded Video Broadcasting Services over LTE-Advanced. \emph{IEEE Trans. Veh. Technol.} \textbf{2015}, \emph{64}, 2186--2192.

\bibitem{TassiJSAC}
Tassi, A.; Chatzigeorgiou, I.; Vukobratovic, D. Resource Allocation Frameworks for Network-coded Layered Multimedia Multicast Services. \emph{IEEE J. Sel. Areas  Commun.} \textbf{2015}, \emph{33}, 141--155.

\bibitem{TassiICC1}
Tassi, A.; Khirallah, C.; Vukobratovic, D.; Chiti, F.; Thompson, J.; Fantacci, R. Reliable rate-optimized video multicasting services over LTE/LTE-A.  In Proceedings of the 2013 IEEE International Conference on Communications (ICC),  Budapest, Hungary, 9--13 June 2013.

\bibitem{TassiICC}
Tassi, A.; Chatzigeorgiou, I.; Vukobratovic, D.; Jones, A. Optimized Network-coded Scalable Video Multicasting over eMBMS Networks.  In Proceedings of the 2015 IEEE International Conference on Communications (ICC),  London, UK, 8--12  June 2015.



\bibitem{5GArchitecture}
3GPP TS 23.501 System Architecture for the 5G System.  Available online: \url{www.3gorg/DynaReport/23501.htm} (accessed on Day Month Year). 

\bibitem{TS38300}
3GPP TS 38.300 NR; Overall Description.  Available online: \url{www.3gorg/DynaReport/38300.htm} (accessed on Day Month Year). 

\bibitem{METIS}
Osseiran, A.; Boccardi, F.; Braun, V.; Kusume, K.; Marsch, P.; Maternia, M.; Queseth, O.; Schellmann, M.; Schotten, H.; Taoka, H.; et al. Scenarios for 5G mobile and wireless communications: the vision of the METIS project.  \emph{IEEE Commun. Mag.} \textbf{2014}, \emph{52}, 26--35.

\bibitem{VideoMEC}
Ge, C.; Wang, N.; Foster, G.; Wilson, M. Toward QoE-Assured 4K Video-on-Demand Delivery Through Mobile Edge Virtualization With Adaptive Prefetching. \emph{IEEE Trans. Multimed.} \textbf{2017}, \emph{19}, 2222--2237.

\bibitem{5GQoS}
Zhang, X.; Cheng, W.; Zhang, H. Heterogeneous statistical QoS provisioning over 5G mobile wireless networks.  \emph{IEEE Netw.}\textbf{ 2014}, \emph{28}, 46--53.





\bibitem{Femtocaching}
Golrezaei, N.; Molisch, A.F.; Dimakis, A.G.; Caire, G.  Femtocaching and device-to-device collaboration: A~new architecture for wireless video distribution. \emph{IEEE Commun. Mag.} \textbf{2013}, \emph{51}, 142--149.


\bibitem{IEEENetwork2017}
Argyriou, A.; Poularakis, K.; Iosifidis, G.; Tassiulas, L. Video Delivery in Dense 5G Cellular Networks. \emph{IEEE~Netw.} \textbf{2017}, \emph{31}, 28--34.

\bibitem{MACaching}
Poularakis, K.; Iosifidis, G.; Sourlas, V.; Tassiulas, L. Exploiting caching and multicast for 5G wireless networks. \emph{IEEE Trans. Wirel. Commun.} \textbf{2016}, \emph{15}, 2995--3007.

\bibitem{HAL}
Borcoci, E.; Ambarus, T.; Bruneau-Queyreix, J.; Negru, D.; Batalla, J.M. Optimization of Multi-server Video Content Streaming in 5G Environment.  In Proceedings of the  International Conference  on Evolving Internet, Barcelona, Spain, 13--17 November 2016.

\bibitem{NCmmWave}
Drago, M.;  Azzino, T.;  Polese, M.; Stefanovic, C.;  Zorzi, M. Reliable Video Streaming over mmWave with Multi Connectivity and Network Coding. \emph{arXiv} \textbf{2017}, arXiv:1711.06154.


\bibitem{5GShen}
Qiao, J.; He, Y.; Shen, X.S. Proactive caching for mobile video streaming in millimeter wave 5G networks.  \emph{IEEE Trans. Wirel. Commun.} \textbf{2016}, \emph{15}, 7187--7198.

\bibitem{URLLC}
Nielsen, J.J.; Liu, R.; Popovski, P. Optimized Interface Diversity for Ultra-Reliable Low Latency Communication (URLLC). \emph{arXiv } \textbf{2017}, arXiv:1712.05148.

\bibitem{PIMRC}
Pervez, F.; Adinoyi, A.; Yanikomeroglu, H. Efficient resource allocation for video streaming for 5G network-to-vehicle communications.  In Proceedings of the 2017 IEEE 28th Annual International Symposium on Personal, Indoor, and Mobile Radio Communications (PIMRC),  Montreal, QC, Canada, \mbox{8--13 October 2017}.



\end{thebibliography}

\end{document}